\documentclass[12pt]{article}

\usepackage{amsmath,amsfonts,amssymb,latexsym}
\usepackage{euscript}
\usepackage{hhline}
\usepackage{graphicx}

\usepackage[all]{xy}
\xyoption{arc} 

\usepackage{epstopdf}

\allowdisplaybreaks[1]

\newlength{\fighskip} \fighskip=2pt
\newlength{\figvskip} \figvskip=3pt

\newcommand*{\figbox}[2]{{
  \def\figscale{#1}
  \def\arraystretch{0.8}
  \arraycolsep=0pt
  \begin{array}{c}
    \vbox{\vskip\figscale\figvskip
      \hbox{\hskip\figscale\fighskip
        \includegraphics[scale=\figscale]{#2}}}
  \end{array}}}

 \usepackage{sectsty}
 \sectionfont{\large}

\newcounter{thm}

\newtheorem{theorem}[thm]{Theorem}
\newtheorem{definition}[thm]{Definition}

\newtheorem{remark}[thm]{Remark}
\newtheorem{lemma}[thm]{Lemma}
\newtheorem{conjecture}[thm]{Conjecture}

\begin{document}

\begin{center}
{\large \bf Some universal properties of Levin-Wen models}
\end{center}

\vspace{0cm}

\begin{center}
  {\normalsize Liang Kong \footnote{Email: kong.fan.liang@gmail.com}}\\
  {\normalsize\it Institute for Advanced Study, Tsinghua University, Beijing, 100084, China}
\end{center}


\vspace{0cm}

\begin{abstract}
We review the key steps of the construction of Levin-Wen type of models on lattices with boundaries and defects of codimension 1,2,3 in a joint work with Alexei Kitaev \cite{kk}. We emphasize some universal properties, such as boundary-bulk duality and duality-defect correspondence, shared by all these models. New results include a detailed analysis of the local properties of a boundary excitation and a conjecture on the functoriality of the monoidal center. 
\end{abstract}



\section{Introduction} \label{sec:intro}
Levin and Wen introduced their lattice models \cite{lw-mod}, each of which is built on the data of a unitary tensor category $\EuScript{C}$, to describe a large class of topological orders \cite{wen}. 
In a joint work with Alexei Kitaev \cite{kk}, we defined and studied Levin-Wen type of models on lattices with boundaries and defects of codimension 1,2,3 by applying the representation theory of tensor categories. 
We also characterized the boundary excitations as module functors. The characterization of the bulk excitations and excitations on a domain wall (or a defect line) can be obtained as a special case by using the so-called ``folding trick". In this article, we will briefly review our construction. Instead of repeating everything, we choose to explain in details a few subtle points in our constructions, including a detailed analysis of the local properties of a boundary excitation, and a few important consequences, some of which are not explicitly stated before. In particular, we will emphasize two universal properties shared by all Levin-Wen models with gapped boundaries and defects: 
\begin{enumerate}
\item {\it boundary-bulk duality}:  A boundary theory, as a system of quasi-particles living on a boundary, determines the bulk theory uniquely by taking center; a bulk theory determines its boundary theories uniquely up to Morita equivalence. 
\item {\it duality-defect correspondence}: There is a canonical group isomorphism between the automorphism group of a bulk theory and the Picard group consisting of invertible domain walls.
\end{enumerate}
These two properties of Levin-Wen models can be extended to a conjecture on the {\it functoriality of $Z$} (see Section \ref{sec:defect}). These results are not isolated phenomena. In 2-dimensional rational conformal field theories (RCFTs), a de-categorified version of above results were proved \cite{ffrs,kr,dkr1,dkr2,dkr3}. Meanwhile, our construction can also be viewed as a physical realization of the so-called extended Turaev-Viro TQFTs \cite{TV,bw,fhlt,dss}. We believe that these results also hold in other extended TQFTs \cite{freed,baez-dolan,lurie}.

\section{The original Levin-Wen models}
\begin{figure}
 \raisebox{-100pt}{
  \begin{picture}(160,150)
   \put(160,8){\scalebox{1.8}{\includegraphics{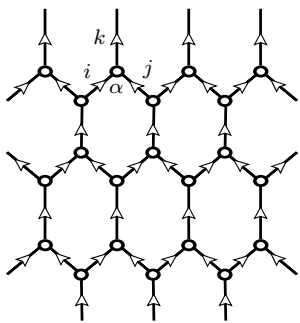}}}
   \put(160,8){
     \setlength{\unitlength}{.75pt}\put(-151,-235){
     \put(190,360)  {\scriptsize $ i $}
     \put(220,362)  {\scriptsize $ j $}
     \put(195,377)  {\scriptsize $ k $}
     \put(203, 352) {\scriptsize $\alpha$}
     }\setlength{\unitlength}{1pt}}
  \end{picture}}
\caption{\label{fig:lw-mod} An upward-oriented planar graph with edge and vertex labels.}
\end{figure}

Let $\EuScript{C}$ be a unitary $\mathbb{C}$-linear finite spherical fusion category. We will briefly recalled the crucial ingredients of such a structure to set the notation. Such category $\EuScript{C}$ has finite inequivalent simple objects $i, j, k, \cdots, m, n\in I$ where $I$ is a finite set, and every object $X$ is a direct sum of simple objects. The tensor product $X \otimes Y$ is well-defined and is determined by the information of the hom spaces $V_i^{XY}:=\text{Hom}_\EuScript{C}(i, X \otimes Y), \forall i\in I$ which are vector spaces over $\mathbb{C}$. In particular, $N_k^{ij} : = \dim V_k^{ij}$ is called the fusion rule. The associator: $(X \otimes Y) \otimes Z \xrightarrow{\alpha_{X,Y,Z}} X \otimes (Y\otimes Z)$ is an isomorphism which satisfies the pentagon relations. It induces an isomorphism between the hom spaces: 
$F_{\,n}^{jim}: \oplus_{k} V_{k}^{ji}\otimes V_{n}^{km} \to \oplus_{l}V_{n}^{jl}\otimes V_{l}^{im}$
which can be expressed by fusion matrices with a choice of basis of hom spaces as follows:
$$
\figbox{1.0}{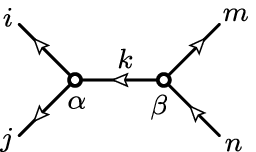} \quad=\,\,
\sum_{l}\sum_{\mu,\nu}\,
\langle l,\mu,\nu|F_{\,n}^{jim}|k,\alpha,\beta\rangle \,\,\,
\figbox{1.0}{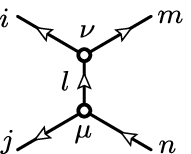}\quad.
$$
The corresponding transformation of graphs is called an ``$F$-move''. The pentagon relations of the associator give the pentagon identities of the fusion matrices. The tensor unit $\textbf{1}$ is an object in $\EuScript{C}$ such that there are the unit isomorphisms $\textbf{1} \otimes X \simeq X \simeq X\otimes \textbf{1}$ satisfying triangle relations. 
Each object $X$ has a two-sided dual $\overline{X}$ satisfying duality axioms so that $\EuScript{C}$ is a spherical fusion category\footnote{It allows one to play on the graphs with the so-called isotopic calculus \cite{kitaev06}, which will not be used here. We will simply choose all edges in our lattice models to oriented upwards (see Fig. \ref{fig:lw-mod}).}.

By the unitarity of $\EuScript{C}$, there is a positive definite Hermitian form $(\cdot, \cdot): \text{Hom}_\EuScript{C}(k, i\otimes j) \otimes \text{Hom}_\EuScript{C}(k, i\otimes j) \to \mathbb{C}$ defined as follows: 
\begin{equation}\label{eq:inner_product}
\langle\eta|\xi\rangle\,=\,
\frac{1}{\sqrt{d_id_jd_k}}\,\,\figbox{1.0}{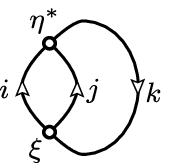}\qquad\quad
\text{for}\,\: \xi,\eta\in V^{ij}_{k},
\end{equation}
where $\eta^*\in V^{k}_{ij}$ is the adjoint morphism, and $d_X$ is the quantum dimension of $X$. By choosing an orthonormal basis $\{ \alpha \}$ of $V_k^{ij}$, we can decompose the identity isomorphism $\text{id}_{i\otimes j}$ as follows:
\begin{equation}\label{eq:decomp_id}
\figbox{1.0}{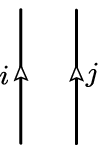}\,=\,\sum_{k\in I}\sum_{\alpha}
\sqrt{\frac{d_k}{d_id_j}}\,\figbox{1.0}{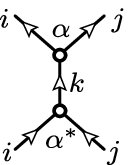}\,\, .
\end{equation}

\bigskip
Now we are ready to define Levin-Wen models. A ``lattice" (or a ``string-net" by Levin and Wen) on a plane is an upward-oriented planar graph with or without external edges. The total Hilbert space $\EuScript{H}_{\text{tot}}$ is defined by $\EuScript{H}_{\text{tot}}:=\otimes_\mathbf{v} \EuScript{H}_\mathbf{v}$ where $\EuScript{H}_\mathbf{v}$ is the space of spins assigned to the vertex $\mathbf{v}$. Let $G$ be any sublattice. We define $\EuScript{H}_G := \otimes_{\mathbf{v} \in G} \EuScript{H}_\mathbf{v}$.  
A \emph{spin} is given by a orthonormal basis vector in an associated morphism space. For example, the spin $\alpha_{ij}^k$ in Fig. \ref{fig:lw-mod} is a basis vector of $\text{Hom}_\EuScript{C}(i\otimes j, k)$ as a subspace of $\EuScript{H}_\mathbf{v}=\oplus_{i,j,k}\text{Hom}_\EuScript{C}(i\otimes j, k)$. 
There is an important subspace $\EuScript{L} \subset \EuScript{H}_{\text{tot}}$. Two spins connected by an internal edge is called compatible if their indices associated to this internal edge coincide. The subspace $\EuScript{L}$ is spanned by compatible spin configurations. We will denote the space of compatible spin configurations in $\EuScript{H}_G$ for any sub-lattice $G$ by $\EuScript{L}_G$. 

\begin{figure}[t]
\centerline{\begin{tabular}{@{}c@{\qquad}c@{\qquad}c@{\qquad}c@{}}
\scalebox{0.85}{\includegraphics{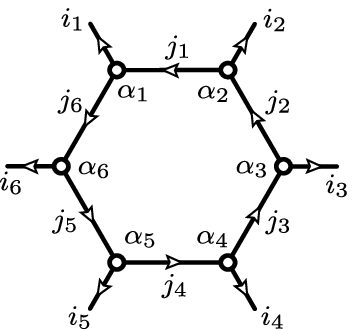}}&
\scalebox{0.85}{\includegraphics{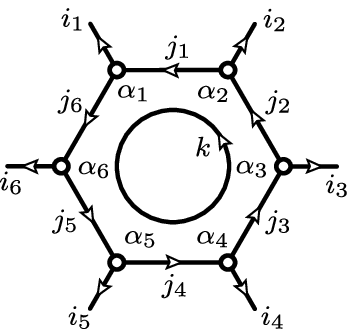}}&
\scalebox{0.85}{\includegraphics{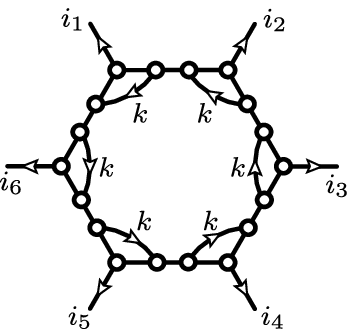}}&
\scalebox{0.85}{\includegraphics{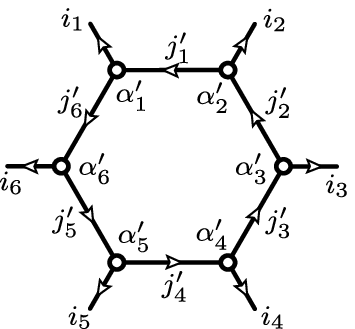}}\\[5pt]
a) & b) & c) & d)
\end{tabular}}
\caption{The action of the plaquette operator $B_{\mathbf{p}}^k$:\newline
a) the initial state of the plaquette;\quad
b) a symbolic representation of the action of $B_{\mathbf{p}}^k$;
c) the loop is partially fused using Eq.~(\ref{eq:decomp_id}) (some labels and the overall factor are not shown);
d) the corner triangles have been evaluated to trivalent vertices (summation over $j_p'$, $\alpha_q'$ is assumed).}\label{fig:Bpk}
\end{figure}

The Hamiltonian of the Levin-Wen model $H:=H_0+H_1$ can be understood as a two-step construction:
\begin{enumerate}
\item The Hamiltonian $H_0 = \sum_{\mathbf{e}} (1- Q_{\mathbf{e}})$, where the sum is over all internal edges $\mathbf{e}$, and $Q_{\mathbf{e}}$ acts as identity if two connected spins are compatible, and as zero if otherwise.  In other words, $\prod_{\mathbf{e}} Q_{\mathbf{e}}$ projects $\EuScript{H}_{\text{tot}}$ onto $\EuScript{L}$. 

\item The additional term $H_1=\sum_{\mathbf{p}} (1- B_{\mathbf{p}})$, where the sum is over all plaquettes $\mathbf{p}$ and $B_{\mathbf{p}}$ acts on the plaquette $\mathbf{p}$ as a projector, defines the ground subspace inside $\EuScript{L}$. More precisely, $B_{\mathbf{p}}$ acts as zero if plaquette $\mathbf{p}$ contain un-compatible spins, and acts as 
\begin{equation}
B_{\mathbf{p}}=\sum_{k\in I}\frac{d_k}{D^2}B_{\mathbf{p}}^k,\quad D^2=\sum_{i\in I} d_i^2.
\end{equation}
where $B_{\mathbf{p}}^k$ is defined in Fig. \ref{fig:Bpk}, if otherwise. 
\end{enumerate}

The operator $B_{\mathbf{p}}$ looks quite mysterious. Notice that $[Q_{\mathbf{v}}, B_{\mathbf{p}}]=0$ for all $\mathbf{v}$ and $\mathbf{p}$. Therefore, the ground state is defined by the following stabilizer conditions: 
\begin{equation}
Q_{\mathbf{v}}|\psi\rangle=|\psi\rangle,\quad\,\,
B_{\mathbf{p}}|\psi\rangle=|\psi\rangle\qquad\,\,
\text{for all $\mathbf{v}$ and $\mathbf{p}$.}
\end{equation}
Therefore, $B_{\mathbf{p}}$ is nothing but the projector from $\EuScript{L}$ to the space of ground states $V$. For Levin-Wen models defined on a plane,
$V$ is nothing but the Hom space defined by the external legs of the graph. A vector in $\EuScript{L}$ can be viewed as a graphic expression of a composed morphism valued in $V$. This gives a map $\text{eval}: \EuScript{L} \to V$. Let $p$ be the number of plaquettes in the graph. One can show that $\text{ev}^\dagger:=D^{-p}\text{eval}^\dagger$ gives an isometric embedding $V \hookrightarrow \EuScript{L}$. Then, one can prove that
$$
\text{ev} \cdot \text{ev}^\dagger=1, \quad\quad \text{ev}^\dagger \cdot \text{ev} = \prod_{\mathbf{p}} B_{\mathbf{p}}.
$$
One can certainly define the Hamiltonian in terms of $\text{ev}$ and $\text{ev}^\dagger$ for simplicity. 
The convenience of using $B_{\mathbf{p}}$ operator is justified later when we study excitations.

\section{Boundary excitations}
\begin{figure}  \label{fig:b-excitation}
\centerline{\begin{tabular}{@{}c@{\qquad\quad}c@{\qquad}c@{\qquad}c@{}}
\scalebox{0.78}{\includegraphics{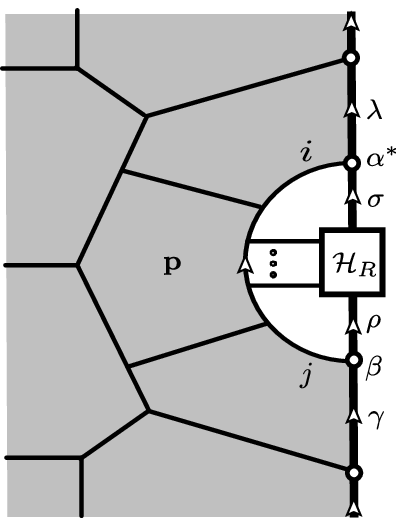}}&
\scalebox{0.78}{\includegraphics{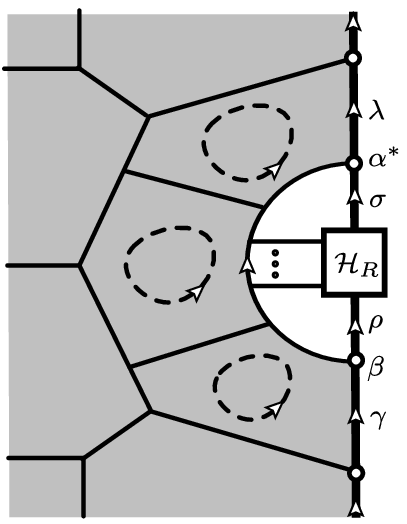}}&
\scalebox{0.78}{\includegraphics{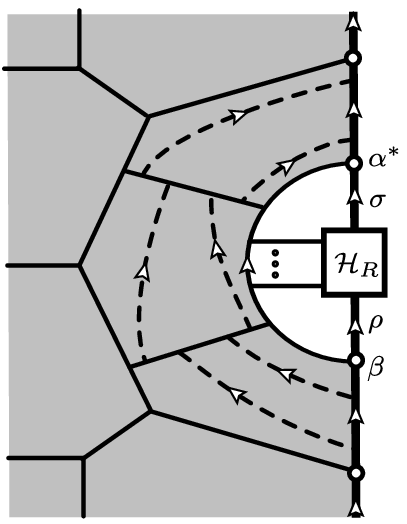}}&
\scalebox{0.78}{\includegraphics{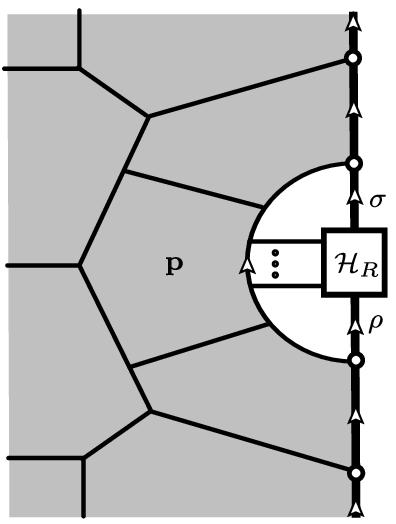}}\\[5pt]
a) & b) & c) & d)
\end{tabular}}
\caption{Boundary excitations (the unexcited part of the lattice is shown in gray):\newline
a) a boundary excitation is localized in the region $R$ and $R'=R \cup \partial R'$ where the lattice configuration at the boundary of $R'$ is $\partial R'=G^{(2,n)}$ (see equation (\ref{eq:G}));\quad
b) three $B_{\mathbf{p}}$operators act on the adjacent plaquette;
c) the loop is partially fused in each plaquette so that it splits into two parts, one of which acts on $\EuScript{H}_R$, the other acts on $\EuScript{H}_{\text{ext}}$;\quad d) the loop is completely fused.}\label{fig:edge-alg-action}
\end{figure}

The construction of Levin-Wen model can be extended to a lattice with a boundary defined by an indecomposable  semisimple $\mathbb{C}$-linear unitary module category ${}_\EuScript{C} \EuScript{M}$ over $\EuScript{C}$, a notion which was introduced by Ostrik \cite{ostrik}. For example, the boundary spin $\alpha^\ast$ in Fig. \ref{fig:edge-alg-action} is a basis vector in $\text{Hom}_\EuScript{M}(i\otimes \sigma, \lambda)$. We will not repeat all steps. Instead, we will choose to comment on a few crucial points and their subtleties. The first crucial point is the definition of an excitation on an ${}_\EuScript{C} \EuScript{M}$-boundary.

\begin{definition}  {\rm
A $\EuScript{M}$-boundary excitation is defined by a pair $(\EuScript{H}_R, P_R)$ where 
\begin{enumerate}
\item $R$ is a region near the lattice boundary (see Fig. \ref{fig:b-excitation}) such that some of the stabilizer conditions are broken or altered within this region;
\item $\EuScript{H}_R$ is the Hilbert space associated to $R$. Namely, $\EuScript{H}_R$ is spanned by all decorations of the region $R$ by spin labels and edge labels together with possible additional extra degrees of freedom that are not present in the original model. 
\item $P_R: \EuScript{H}_R \to \EuScript{H}_R$ is a projector such that $P_R$ commutes with the partial action of the plaquette operator $B_{\mathbf{p}}$ outside the region $R$. More precisely, the operator $B_{\mathbf{p}}$ can be expressed as $\sum_q Y_q \otimes X_q$ where $X_q$ acts on $\EuScript{H}_R$ and $Y_q$ acts on $\EuScript{H}_{\text{ext}}$ which is the Hilbert space associated to the region outside of $R$. Assuming that $Y_q$ are linear independent, we will call $X_q$ {\it partial plaquette operators}. Then $P_R$ is required to satisfy the condition $[P_R, X_q]=0$ for all $q$. 
\end{enumerate}
}
\end{definition}
As an immediate consequence, an excitation is equivalent to a module over the local operator algebra generated by partial plaquette operators. But there is an ambiguity in this definition since the choice of region $R$ is quite arbitrary.

The first solution to this ambiguity is to reduce the general cases to the simplest case \cite{kk}. More precisely, one can always alter the lattice by ``F-moves" and ``moving-bubbles" to simplify any boundary configuration $\partial R$ of a lattice region $R$ to $G^{(0,0)}$, which is defined in equation (\ref{eq:G}), without changing the physics.  In particular, the Hamiltonian is preserved under these lattice transformations.

The second solution is to study directly the relation between two general choices of $R$. In order to formulate the relation more precisely, we introduce the notion of morphism between two excitations $(\EuScript{H}_R, P_R)$ and $(\EuScript{H}_R, Q_R)$. It is defined as a linear map $\EuScript{H}_R \to \EuScript{H}_R$ such that it commutes with the partial plaquette operators. Such defined excitations and morphisms between them form a category $\EuScript{E}_R$. We would like to show that $\EuScript{E}_R \cong \EuScript{E}_{R'}$ for different choice of regions $R$ and $R'$. It is enough to consider the cases in which $R'$ contains $R$ in a minimal but nontrivial way. Let $R \subset R'$ where $R'$ is the region containing both the region $R$ and a boundary diagram $G^{(m,n)}=\partial R'$ of the following type: 
\begin{equation}  \label{eq:G}
G^{(m,n)} = \raisebox{-30pt}{
 \begin{picture}(120,70)
   \put(10,-5){\scalebox{1}{\includegraphics{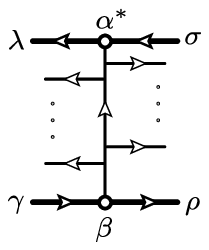}}}
   \put(10, 40){ \tiny $1$}
   \put(8, 18){ \tiny $m$}
   \put(56,45){ \tiny $1$}
   \put(56,22){ \tiny $n$}
 \end{picture}
 }
\end{equation}
In part a) of Fig. \ref{fig:b-excitation}, we give an example of such situation when $m=2$. We have $\EuScript{H}_{R'}= \EuScript{H}_{\partial R'} \otimes \EuScript{H}_R$. We denote the region outside $R'$ by $\text{ext}$. Then we have $\EuScript{H}_{\text{tot}} = \EuScript{H}_{\text{ext}} \otimes \EuScript{H}_{R'}$. 
Part b) in Fig. \ref{fig:b-excitation} portraits the action of 3 plaquette operators $B_{\mathbf{p}}$ immediately adjacent to the region $R'$. Part c) portraits how the action splits into two parts with the right dash lines, which are the partial plaquette operators in this case, acting on $\EuScript{H}_{R'}$ and the left dash lines acting on $\EuScript{H}_{\text{ext}}$ by the definition of $B_{\mathbf{p}}$ (recall Fig. \ref{fig:Bpk}). The partial plaquette operators span the space $A^{(m,m)}$, which is nothing but $\EuScript{L}_{G^{(m,m)}}$ and is closed under the multiplication\footnote{The definition of multiplication is similar to the case $m=0=n$ which is defined explicitly in \cite[eq.(18)]{kk}.} of the partial plaquette operators. Namely, $A^{(m,m)}$ is an algebra.
By definition, we have $[P_{R'}, A^{(m,m)}]=0$. In other words, $(\EuScript{H}_{R'}, P_{R'})$ defines a left $A^{(m,m)}$-module $\text{Im} P_{R'}$. Moreover, morphisms between excitations are precisely left $A^{(m,m)}$-module maps. 
We obtain that $\EuScript{E}_{R'}$ is equivalent to the category of left $A^{(m,m)}$-modules. Similarly, $\EuScript{E}_R$ is equivalent to the category of left $A^{(n,n)}$-modules. Notice that the space $A^{(m,n)}:=\EuScript{L}_{\partial R'}$ carries a natural structure of $A^{(m,m)}$-$A^{(n,n)}$-bimodules. 

\begin{lemma} \label{lem:morita}  
The algebras $A^{(m,m)}$ and $A^{(n,n)}$ are Morita equivalent for any $m,n\in \mathbb{N}$. Moreover, the $A^{(m,m)}$-$A^{(n,n)}$-bimodule $A^{(m,n)}$ is invertible and defines the Morita equivalence. 
\end{lemma}

In other words, the boundary graph $G^{(m,n)}$ exactly determines an equivalence $\EuScript{E}_R \to \EuScript{E}_{R'}$ between the categories of $\EuScript{M}$-boundary excitations defined by $R$ and $R'$, respectively. Translating the invertible bimodule $A^{(m,n)}$ into the defining language of an excitation, this equivalence is given by a functor:
$$
(H_R, P_R) \mapsto (H_{R'},  \prod_{\mathbf{p} \in Q} B_{\mathbf{p}} \cdot P_R) 
$$
where $Q$ is the set of plaquettes sitting between the region $R$ and $\partial R'$. 
More details of above discussion will be covered in our future publication \cite{kong}.

The Morita equivalence between $A^{(m,m)}$ and $A^{(n,n)}$ allows us to identify the category of $\EuScript{M}$-boundary excitations with the category of modules over $A^{(0,0)}$ which actually carries a structure of weak $C^\ast$-Hopf algebra \cite{kk}. 
One can further show that the category of $A^{(0,0)}$-modules is equivalent to the category $\text{Fun}_\EuScript{C}(\EuScript{M}, \EuScript{M})$ of $\EuScript{C}$-module functors \cite{ostrik,kk}. We use 
$\EuScript{C}_\EuScript{M}^\vee$ to denote the category $\text{Fun}_\EuScript{C}(\EuScript{M}, \EuScript{M})^{\otimes^{\mathrm{op}}}$, which is the same category as $\text{Fun}_\EuScript{C}(\EuScript{M}, \EuScript{M})$ but with the tensor product $\otimes^{\mathrm{op}}$ defined by $F \otimes^{\mathrm{op}} G := F\otimes G$ for $F,G\in \text{Fun}_\EuScript{C}(\EuScript{M}, \EuScript{M})$.
We obtain our first fundamental result:

\begin{theorem}
Excitations on an $\EuScript{M}$-boundary in Levin-Wen models are classified by the objects in the category $\EuScript{C}_\EuScript{M}^\vee$, which is also a unitary finite fusion category. 
\end{theorem}

\begin{remark}   {\rm
Although anyons in physics often refer to simple objects in a semisimple braided tensor category, objects in a unitary fusion category without braiding are quasiparticles that can only live on a line like the boundary line in our case. Moreover, the direct sums of simple objects are physically unavoidable because the tensor product of two simple objects can split into a direct sum of simples objects in general. For example, in the Ising model, we have $\sigma \otimes \sigma = 1 \oplus \epsilon$,
which means that the fusion of two $\sigma$ anyons can split into either $1$ or $\epsilon$. This should be viewed as a categorification of the superposition of pure states in Quantum Mechanics. Therefore, an anyon associated to a simple object can be called a {\it pure anyon}. A general anyon is {\it a superposition of pure anyons}, and can be called a {\it mixed anyon}. 
}
\end{remark}

Note that $\EuScript{C}_\EuScript{C}^\vee \simeq \EuScript{C}$ as tensor categories. In general, different boundaries $\EuScript{M} \ncong \EuScript{N}$ give arise to different boundary excitations $\EuScript{C}_\EuScript{M}^\vee$ and $\EuScript{C}_\EuScript{N}^\vee$, which are, nevertheless, Morita equivalent. They share the same bulk theory (see Section \ref{sec:defect}) given by their monoidal centers, i.e. $Z(\EuScript{C}_\EuScript{M}^\vee) \cong Z(\EuScript{C}) \cong Z(\EuScript{C}_\EuScript{N}^\vee)$ \cite{mueger1,eno}. 
This is due to the fact that $\EuScript{C}$ and $\EuScript{D}$ are Morita equivalent iff $Z(\EuScript{C}) \cong Z(\EuScript{D})$ as braided tensor categories \cite{eno,kitaev}.  
Conversely, any unitary finite fusion category $\EuScript{D}$, which is Morita equivalent to $\EuScript{C}$, can be realized as a boundary theory of the same $\EuScript{C}$-bulk lattice model. Indeed, let $\EuScript{M}$ be an invertible $\EuScript{C}$-$\EuScript{D}$-bimodule that defines the Morita equivalence between $\EuScript{C}$ and $\EuScript{D}$. Then the excitations of an ${}_\EuScript{C} \EuScript{M}$-boundary is given by $\EuScript{D}$. These results are nothing but the {\it boundary-bulk duality} mentioned in Section \ref{sec:intro}.

\section{Defects of codimension 1,2,3}\label{sec:defect}
The theory of domain wall (or defect line) can be obtained from the boundary theory by using the folding trick. Indeed, an ${}_\EuScript{C}\EuScript{M}_\EuScript{D}$-wall can be equivalently viewed as an ${}_{\EuScript{C} \boxtimes \EuScript{D}^{\otimes^{\text{op}}}} \EuScript{M}$-boundary by folding the system along the $\EuScript{M}$-wall. Therefore, we obtain
\begin{theorem}
Excitations on an ${}_\EuScript{C}\EuScript{M}_\EuScript{D}$-wall are classified by the objects in the category $\text{Fun}_{\EuScript{C}|\EuScript{D}}(\EuScript{M}, \EuScript{M})$ of $\EuScript{C}$-$\EuScript{D}$-bimodule functors. The bulk excitations, which can be viewed as excitations on a ${}_\EuScript{C}\EuScript{C}_\EuScript{C}$-wall,  are classified by the objects in the monoidal center $Z(\EuScript{C}):=\text{Fun}_{\EuScript{C}|\EuScript{C}}(\EuScript{C}, \EuScript{C})$ which is actually a modular tensor category \cite{mueger2}. 
\end{theorem}
Both $\EuScript{C}$-bulk excitations and $\EuScript{D}$-bulk excitations can be moved onto an ${}_\EuScript{C}\EuScript{M}_\EuScript{D}$-wall and become wall excitations via the following monoidal functors:  
$$
Z(\EuScript{C}) \xrightarrow{L_\EuScript{M}} \text{Fun}_{\EuScript{C}|\EuScript{D}}(\EuScript{M}, \EuScript{M}), \quad\quad \text{Fun}_{\EuScript{C}|\EuScript{D}}(\EuScript{M}, \EuScript{M}) \xleftarrow{R_\EuScript{M}} Z(\EuScript{D}),
$$
which are defined by
\begin{eqnarray}
L_\EuScript{M}:  (\EuScript{C} \xrightarrow{\EuScript{F}} \EuScript{C})  &\mapsto& (\EuScript{M} \simeq \EuScript{C}\boxtimes_\EuScript{C} \EuScript{M} \xrightarrow{\EuScript{F} \boxtimes_\EuScript{C} 1} \EuScript{C}\boxtimes_\EuScript{C} \EuScript{M} \simeq \EuScript{M}),  \nonumber \\
R_\EuScript{M}: (\EuScript{D} \xrightarrow{\EuScript{G}} \EuScript{D}) &\mapsto& (\EuScript{M} \simeq \EuScript{M} \boxtimes_\EuScript{D} \EuScript{D} \xrightarrow{1\boxtimes_\EuScript{D} \EuScript{G}}
\EuScript{M} \boxtimes_\EuScript{D} \EuScript{D} \simeq \EuScript{M}). \nonumber
\end{eqnarray}
When $\EuScript{M}$ is invertible, i.e. $\EuScript{C}$ is Morita equivalent to $\EuScript{D}$, $L_\EuScript{M}$ is invertible and its inverse is given by
$$
L_\EuScript{M}^{-1}: (\EuScript{M} \xrightarrow{\EuScript{F}} \EuScript{M}) \mapsto (\EuScript{C} \simeq \EuScript{M}\boxtimes_\EuScript{D} \EuScript{M}^{\text{op}} \xrightarrow{\EuScript{F} \boxtimes_\EuScript{D} 1} \EuScript{M}\boxtimes_\EuScript{D} \EuScript{M}^{\text{op}}\simeq \EuScript{C}),
$$ 
where $\EuScript{M}^{\text{op}}$ is the opposite category of $\EuScript{M}$ and automatically a $\EuScript{C}$-$\EuScript{D}$-bimodule. Similarly, $R_\EuScript{M}$ is also invertible. In this case, $R_\EuScript{M}^{-1} \circ L_\EuScript{M}$, defined by the following conjugate action: 
$$
R_\EuScript{M}^{-1} \circ L_\EuScript{M}: Z(\EuScript{C})=\text{Fun}_{\EuScript{C}|\EuScript{C}}(\EuScript{C}, \EuScript{C}) \to \text{Fun}_{\EuScript{C}|\EuScript{C}}(\EuScript{M}^{\text{op}}\boxtimes_\EuScript{C} \EuScript{C} \boxtimes_\EuScript{C} \EuScript{M}, \EuScript{M}^{\text{op}}\boxtimes_\EuScript{C} \EuScript{C} \boxtimes_\EuScript{C} \EuScript{M}) \cong Z(\EuScript{C})
$$
is actually a braided autoequivalence between $Z(\EuScript{C})$ and $Z(\EuScript{D})$. Therefore, we obtain a map $\EuScript{M} \mapsto R_\EuScript{M}^{-1} \circ L_\EuScript{M}$ between the set of equivalent classes of invertible $\EuScript{C}$-$\EuScript{D}$-bimodules and the set of braided autoequivalences between $Z(\EuScript{C})$ and $Z(\EuScript{D})$. This map turns out to be bijective. When $\EuScript{C}=\EuScript{D}$, we obtain a group isomorphism $Z: \text{Pic}(\EuScript{C}) \xrightarrow{\simeq} \text{Aut}(Z(\EuScript{C}))$ between the Picard group $\text{Pic}(\EuScript{C})$ and the braided auto-equivalence group $\text{Aut}(Z(\EuScript{C}))$ \cite{enom}. In terms of physical language, it is the {\it duality-defect correspondence} mentioned in Section \ref{sec:intro}. A similar duality-defect correspondence holds for RCFTs \cite{dkr1}.

\medskip
A defect junction (or a defect of codimension 2) between an ${}_\EuScript{C}\EuScript{M}_\EuScript{D}$-wall and an ${}_\EuScript{C}\EuScript{N}_\EuScript{D}$-wall can be defined in a way similar to a boundary excitation. An immediate consequence of this definition is that such defect junctions are classified by the modules over certain local operator algebra \cite{kk}, and, equivalently, by $\EuScript{C}$-$\EuScript{D}$-bimodule functor from $\EuScript{M}$ to $\EuScript{N}$. We denote the category of such functors by $\text{Fun}_{\EuScript{C}|\EuScript{D}}(\EuScript{M}, \EuScript{N})$, or simply by $Z(\EuScript{M}, \EuScript{N})$. We also simplify the notation $\text{Fun}_{\EuScript{C}|\EuScript{D}}(\EuScript{M}, \EuScript{M})$ by $Z(\EuScript{M})$. Obviously, a bulk or wall excitation is nothing but a special defect of codimension 2. Notice that $Z(\EuScript{M}, \EuScript{N})$ is naturally a $Z(\EuScript{N})$-$Z(\EuScript{M})$-bimodule. We automatically have a right $Z(\EuScript{M})$-module functor
$\EuScript{F}_\ast: Z(\EuScript{M}) \to Z(\EuScript{M}, \EuScript{N})$ and a left $Z(\EuScript{N})$-module functor
$\EuScript{F}^\ast: Z(\EuScript{N}) \to Z(\EuScript{M}, \EuScript{N})$ defined by 
$$
\EuScript{F}_\ast:  f \mapsto \EuScript{F} \circ f, \,\,\, \forall f\in Z(\EuScript{M}), \quad \quad\quad \EuScript{F}^\ast: g \mapsto g \circ \EuScript{F}, \,\,\, \forall g\in Z(\EuScript{N}),
$$
respectively. Physically, it says that a wall excitation on an $\EuScript{M}$-wall or an $\EuScript{N}$-wall can fuse into a defect junction between the $\EuScript{M}$-wall and the $\EuScript{N}$-wall. This fusion process defines a pair of morphisms $(Z(\EuScript{M}) \xrightarrow{\EuScript{F}_\ast} Z(\EuScript{M}, \EuScript{N}) \xleftarrow{\EuScript{F}^\ast} Z(\EuScript{N}))$, or equivalently a $Z(\EuScript{N})\boxtimes_{\text{Hilb}} Z(\EuScript{M})^{\otimes^{\text{op}}}$-module functor $\EuScript{F}^\ast \boxtimes_{\text{Hilb}} \EuScript{F}_\ast: Z(\EuScript{N})\boxtimes_{\text{Hilb}} Z(\EuScript{M})^{\otimes^{\text{op}}} \to  Z(\EuScript{M}, \EuScript{N})$ defined by $(g, f) \mapsto g\circ \EuScript{F} \circ f$ for $g\in Z(\EuScript{N}), f\in Z(\EuScript{M})$.

\medskip
Defects of codimension 3 are instantons which are given by natural transformations between (bi-)module functors. Let $\EuScript{F}, \EuScript{G}:\EuScript{M} \to \EuScript{N}$ be two $\EuScript{C}$-$\EuScript{D}$-bimodule functors. A natural transformation $\phi: \EuScript{F} \to \EuScript{G}$ induces a natural transformation $Z(\phi)$ between the two $Z(\EuScript{N})\boxtimes_{\text{Hilb}} Z(\EuScript{M})^{\otimes^{\text{op}}}$-module functors: 
$\EuScript{F}^\ast \boxtimes \EuScript{F}_\ast \xrightarrow{Z(\phi)} \EuScript{G}^\ast \boxtimes \EuScript{G}_\ast$. 


\medskip
Combining all previous structures, we obtain two related multilayered structures. 
\begin{enumerate}
\item One is given by the building data of the lattice models:
\begin{equation} \label{diag:lw-data}
\begin{xy} 0;/r.22pc/:
(0,15)*{};
(0,-15)*{};
(0,8)*{}="A";
(0,-8)*{}="B";
{\ar@{=>}@/_1pc/ "A"+(-4,1) ; "B"+(-3,0)_{\EuScript{F}}};
{\ar@{=}@/_1pc/ "A"+(-4,1) ; "B"+(-4,1)};
{\ar@{=>}@/^1pc/ "A"+(4,1) ; "B"+(3,0)^{\EuScript{G}}};
{\ar@{=}@/^1pc/ "A"+(4,1) ; "B"+(4,1)};
{\ar@3{->} (-6,0)*{} ; (6,0)*+{}^{\phi}};
(-15,0)*+{\EuScript{C}}="1";
(15,0)*+{\EuScript{D}}="2";
{\ar@{->}@/^2.75pc/ "1";"2"^{\EuScript{M}}};
{\ar@{->}@/_2.75pc/ "1";"2"_{\EuScript{N}}};
\end{xy} 
\end{equation}
It is natural to conjecture that these four layers of structures together with composition maps give a tricategory \cite{gps}. Equivalently, one can replace $\EuScript{C}$ and $\EuScript{D}$ by the bicategories of the modules\footnote{These modules are required to be unitary, semi-simple and finite.} over them; and $\EuScript{M}, \EuScript{N}$ by corresponding functors $\EuScript{F}_\EuScript{M}, \EuScript{F}_\EuScript{N}$; and $\EuScript{F}, \EuScript{G}$ by corresponding natural transformations $\varphi_\EuScript{F},\varphi_\EuScript{G}$; and $\phi$ by a modification $m_\phi: \varphi_\EuScript{F} \to \varphi_\EuScript{G}$. 
\begin{equation} \label{diag:lw-data-2}
\xy 0;/r.22pc/:
(0,15)*{};
(0,-15)*{};
(0,8)*{}="A";
(0,-8)*{}="B";
{\ar@{=>}@/_1pc/ "A"+(-4,1) ; "B"+(-3,0)_{\varphi_\EuScript{F}}};
{\ar@{=}@/_1pc/ "A"+(-4,1) ; "B"+(-4,1)};
{\ar@{=>}@/^1pc/ "A"+(4,1) ; "B"+(3,0)^{\varphi_\EuScript{G}}};
{\ar@{=}@/^1pc/ "A"+(4,1) ; "B"+(4,1)};
{\ar@3{->} (-6,0)*{} ; (6,0)*+{}^{m_\phi}};
(-15,0)*+{\mbox{\hspace{-1cm}$\EuScript{C}$-Mod}}="1";
(15,0)*+{\mbox{\hspace{+1cm}$\EuScript{D}$-Mod}}="2";
{\ar@/^2.75pc/ "1";"2"^{\EuScript{F}_\EuScript{M}}};
{\ar@/_2.75pc/ "1";"2"_{\EuScript{F}_\EuScript{N}}};
\endxy 
\end{equation}
The bicategory of $\EuScript{C}$-modules is also called {\it the category of boundary conditions} in physics. Indeed, a $\EuScript{C}$-module is precisely a boundary condition for a $\EuScript{C}$-bulk in a Levin-Wen model. 

\item The other multilayered structure is given by the excitations in the models together with various fusion process of excitations from the bulk to defect lines and from defect lines to defect junctions as shown in the following diagram: 
\begin{equation}  \label{diag:excitation}
\xymatrix{
  & & Z(\EuScript{M}) \ar[ldd]_{\EuScript{F}_\ast} \ar[rdd]^{\EuScript{G}_\ast} & & \\
  & & & & \\
Z(\EuScript{C}) \ar[rruu]^{L_\EuScript{M}} \ar[rrdd]_{L_\EuScript{N}}   & Z(\EuScript{M}, \EuScript{N})
      \ar@3{->}[rr]^{Z(\phi)}
  & & Z(\EuScript{M}, \EuScript{N})
    & Z(\EuScript{D}) \ar[lluu]_{R_\EuScript{M}} \ar[lldd]^{R_\EuScript{N}} \\
    & & & &  \\
    & & Z(\EuScript{N}) \ar[luu]^{\EuScript{F}^\ast} \ar[ruu]_{\EuScript{G}^\ast} & & 
}
\end{equation}
Moreover, each of the following subdiagrams: 
$$
\EuScript{F}_\ast \circ L_\EuScript{M} \simeq \EuScript{F}^\ast \circ L_\EuScript{N}, \quad
\EuScript{F}_\ast \circ R_\EuScript{M} \simeq \EuScript{F}^\ast \circ R_\EuScript{N}, \quad
\EuScript{G}_\ast \circ L_\EuScript{M} \simeq \EuScript{G}^\ast \circ L_\EuScript{N}, \quad
\EuScript{G}_\ast \circ R_\EuScript{M} \simeq \EuScript{G}^\ast \circ R_\EuScript{N}, \quad
$$
is commutative up to a canonical isomorphism which is given by the defining data of the 
bimodule functor $\EuScript{F}$  and $\EuScript{G}$.  The pair $(L_\EuScript{M}, R_\EuScript{M})$ is called a cospan. A composition map can be defined for two connected cospans. Similarly, the pair $(\EuScript{F}_\ast, \EuScript{F}^\ast)$ is also a cospan. It is natural to conjecture that above multilayered structure in diagram (\ref{diag:excitation}), together with properly defined composition maps and coherence morphisms, can be embedded into a tricategory, in which the 1- and 2-morphisms are given by cospans. 
\end{enumerate}
The physical intuition seems to suggest that the assignment $Z$ from the building data of the lattice models to physical excitations in the bulks and on the defects is functorial. Namely, we conjecture that $Z$ can be realized as a 3-functor. This functoriality, which will be called the {\it functoriality of $Z$}, is not an isolate phenomenon. It occurs in its simplest form as the fact that the notion of center for an ordinary algebra over a field is a lax functor \cite{dkr2}. A de-categorified version (as a 2-functor) of this conjectured 3-functor is constructed and proved in RCFTs \cite{dkr3}. 

Restricting to the equivalent classes of invertible 1-morphisms in (\ref{diag:lw-data}) and (\ref{diag:excitation}) and ignoring all higher morphisms, it is known that the assignment $Z$ gives a true functor between two groupoids \cite{enom}. In terms of this functor, the boundary-bulk duality and duality-defect correspondence can be easily summarized as the fact that $Z$ is fully-faithful \cite{enom}.

\section{Summary and outlooks}

We have reviewed our construction \cite{kk} of Levin-Wen models enriched by boundaries and defects of codimensions 1,2,3, and emphasized three universal properties: boundary-bulk duality, duality-defect correspondence and the conjectured functoriality of $Z$. We also pointed out that 2-dimensional RCFTs share similar properties. Therefore, it is natural to expect that such properties also hold in more general context. 

Extended TQFTs was invented by mathematicians \cite{freed,baez-dolan,lurie}. Its physical meaning was realized only later \cite{ffrs,bdh,kapustin}. Roughly speaking, an extended $n$-dimensional TQFT can be viewed as an ordinary TQFT (in the sense of Atiyah) enriched by defects of codimension $1$,$2$,...,$n$. For example, the Levin-Wen models with defects provide a physical realization of the extended Turaev-Viro TQFTs \cite{kb,kirillov}; RCFTs with defects can be viewed roughly as an extended 2-d TQFT valued in a nontrivial modular tensor category. It is also known that the finite fusion categories, which are used to define the Levin-Wen models, and its counterparts\footnote{In RCFT, they are special symmetric Frobenius algebras in a modular tensor category \cite{frs}.} in RCFTs are both fully dualizable objects \cite{lurie,fhlt,dss}, which classify the extended TQFTs \cite{baez-dolan}, in the sense of Hopkins-Lurie \cite{lurie}. It is natural to conjecture that these three properties also hold in other extended TQFTs. More precisely, for a fully-dualizable object $\EuScript{C}$, let $Z(\EuScript{C})$ be the Hoshchild cohomology $HH^\bullet(\EuScript{C})$ defined by Lurie \cite{lurie2}, then we propose the following mathematical conjecture: 
\begin{conjecture}
The Hoshchild cohomology $Z(-)$ defines a functor from a category of fully dualizable objects (similar to (\ref{diag:lw-data})) to a target category constructed via a system of cospans similar to what is shown in the diagrams (\ref{diag:excitation}). It is fully faithful if we restrict $Z(-)$ to only $0$-morphisms and the equivalent classes of invertible $1$-morphisms, ignoring higher morphisms, in the domain category. In particular, it says that, for two fully dualizable objects $\EuScript{C}$ and $\EuScript{D}$, $Z(\EuScript{C})$ is isomorphic (as for example $E_n$-categories, depending on the context) to $Z(\EuScript{D})$, if and only if $\EuScript{C}$ and $\EuScript{D}$ are Morita equivalent; and we have a group isomorphism $\text{Pic}(\EuScript{C}) \simeq \text{Aut}(Z(\EuScript{C}))$.
\end{conjecture} 

It is possible to generalize Levin-Wen models in different ways other than increasing the dimension. 
Note that the construction of Levin-Wen models is not sensitive to the condition that the tensor unit $\textbf{1}$ in $\EuScript{C}$ is simple. The same construction extends to multi-fusion categories. If $\EuScript{C}$ is a non-trivial multi-fusion category, it is possible that $Z(\EuScript{C})\cong \text{Vect}_\mathbb{C}$ \cite{enom} which describes a trivial topological phase. One can also define a lattice model associated to a unitary spherical fusion category $\EuScript{C}$ equipped with a fiber functor $\omega: \EuScript{C} \to \text{Hilb}$. Such models can reproduce Kitaev's quantum double models associated to a finite group \cite{kitaev97} or a $C^\ast$-Hopf algebra \cite{bmca}. It will be interesting to study boundaries and defects in these models \cite{bs}.

\section*{Acknowledgments}
I would like to thank Alexei Kitaev for a very productive collaboration, and for his generosity of sharing many of his unpublished works with me, and for his support in my career. I want to thank Zheng-Han Wang, Xiao-Gang Wen, Yong-Shi Wu and Yi-Zhuang You for their valuable comments. I am also very grateful for the session chairs, Karl-Henning Rehren and Jean-Michel Maillet, of ICMP 2012 for inviting me to speak in this conference. This work is supported by the Basic Research Young Scholars Program and the Initiative Scientific Research Program of Tsinghua University, and NSFC under Grant No. 11071134.


\end{document}